 \documentclass[aps,pra,twocolumn,superscriptaddress]{revtex4-1}

\usepackage{amsthm}
\usepackage{amsmath,bm}
\usepackage{amssymb}
\usepackage{amsfonts}
\usepackage{graphicx}

\usepackage{txfonts}

\usepackage{float}
\usepackage{quantum} 

\usepackage[colorlinks=true,linkcolor=blue,citecolor=blue,urlcolor=blue]{hyperref}

\begin{document}

\newcommand{\avg}[1]{\langle#1\rangle}		
\newcommand{\var}{\text{Var}}	
\newcommand{\mse}{\text{MSE}}	
\newcommand{\cov}{\text{Cov}}

\newcommand{\Vs}[1]{\var[#1]}
\newcommand{\MSE}[1]{\mse[#1]}
\newcommand{\cvar}{\var_Q^{B|A}}

\newcommand{\ie}{\textit{i.e.\ }}

\renewcommand{\vec}[1]{\mathbf{#1}}
\newcommand{\bg}[1]{\boldsymbol{#1}}
\newcommand{\inlineheading}[1]{\textit{{#1.---}}}
\newcommand{\mc}[1]{\mathcal{#1}}
\newcommand{\mb}[1]{\mathbf{#1}}
\newcommand{\comment}[1]{{\color{red}[#1]}}
\newcommand{\fisher}{F}
\newcommand{\cfi}{\fisher^{B|A}}
\newcommand{\qfi}{\fisher_Q}
\newcommand{\cqfi}{\fisher_Q^{B|A}}

\newcommand{\assem}{\mc{A}}
\newcommand{\tqfi}{\bar{\mc{F}}}
\newcommand{\tcqfi}{\bar{\mc{F}}^{B|A}}
\newcommand{\steer}{\mc{S}}
\newcommand{\steerAvg}{\steer_\mathrm{avg}}
\newcommand{\steerMax}{\steer_\mathrm{max}}
\newcommand{\pos}[1]{\left[ #1 \right]^+}

\newcommand{\ThAP}{\theta_A^\prime}

\newcommand{\mg}[1]{{\color{magenta}[(MG) #1]}}
\newcommand{\ps}[1]{{\color{red}[PS:  #1]}}
\newcommand{\mf}[1]{{\color{blue}[(MF) #1]}}
\newcommand{\jj}[1]{{\color{cyan}[(JJ) #1]}}

\newcommand{\new}[1]{{\color{red}#1}}

\newcommand{\tap}{\theta_A^\prime}

\title{Assisted metrology and preparation of macroscopic superpositions\\ with split spin-squeezed states}

\author{Jiajie Guo}
\affiliation{State Key Laboratory for Mesoscopic Physics, School of Physics, Frontiers Science Center for Nano-optoelectronics, $\&$ Collaborative
Innovation Center of Quantum Matter, Peking University, Beijing 100871, China}

\author{Fengxiao Sun}
\affiliation{State Key Laboratory for Mesoscopic Physics, School of Physics, Frontiers Science Center for Nano-optoelectronics, $\&$ Collaborative
Innovation Center of Quantum Matter, Peking University, Beijing 100871, China}

\author{Qiongyi He}
\email{qiongyihe@pku.edu.cn}
\affiliation{State Key Laboratory for Mesoscopic Physics, School of Physics, Frontiers Science Center for Nano-optoelectronics, $\&$ Collaborative
Innovation Center of Quantum Matter, Peking University, Beijing 100871, China}
\affiliation{Collaborative Innovation Center of Extreme Optics, Shanxi University, Taiyuan, Shanxi 030006, China}
\affiliation{Peking University Yangtze Delta Institute of Optoelectronics, Nantong 226010, Jiangsu, China}

\author{Matteo Fadel}
\email{fadelm@phys.eth.ch}
\affiliation{Department of Physics, ETH Z\"{u}rich, 8093 Z\"{u}rich, Switzerland}

\date{\today}

\begin{abstract}
    We analyse the conditional states in which one part of a split spin-squeezed state is left, upon performing a collective spin measurement on the other part. For appropriate measurement directions and outcomes, we see the possibility of obtaining states with high quantum Fisher information, even reaching the Heisenberg limit. This allows us to propose a metrological protocol that can outperform standard approaches, for example in a situation where the number of particles in the probe is bounded. The robustness of this protocol is investigated by considering realistic forms of noise present in cold-atom experiments, such as particle number fluctuations and imperfect detection. Ultimately, we show how this measurement-based state preparation approach can allow for the conditional (\ie heralded) preparation of spin Schr\"{o}dinger's cat states even when the initial state before splitting is only mildly squeezed.
\end{abstract}

{
\let\clearpage\relax
\maketitle
}

\begin{figure}[t]
    \begin{center}
	\includegraphics[width=80mm]{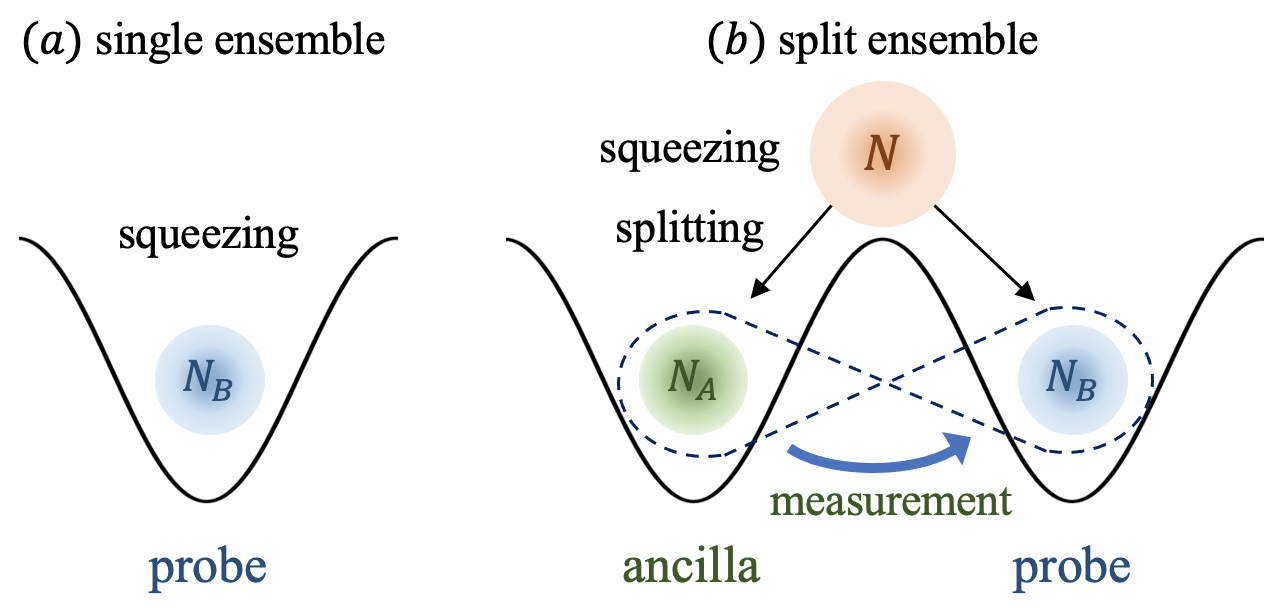}
	\end{center}
    \caption{\textbf{State preparation and metrology with split atomic ensembles.} (a) To prepare non-classical states of many-body systems, traditional approaches rely on implementing a nonlinear dynamic in a trapped ensemble. (b) Instead, our assisted protocol is based on first spatially splitting a mildly entangled ensemble, to then measure one of the two parts. Because of the shared correlations, this results in projecting the other half into a multipartite state that can have strong quantum correlations.}
    \label{Fig1}
\end{figure}

\section{Introduction}
Spin-squeezed states are of paramount importance for investigating multipartite quantum correlation, as well as for quantum-enhanced metrology applications. Experimentally, these states are nowadays routinely prepared in atomic ensembles, either by controlling atomic collisions, or by light-matter interaction. In these platforms, a number of studies revealed the rich entanglement structure of spin-squeezed states \cite{FrerotArXiv2023}, and demonstrated their usefulness for performing measurements with a precision surpassing the standard quantum limit \cite{PezzeRMP2018}.

Recently, the concept of split spin-squeezed states was introduced, where an ensemble of spin-squeezed particles is spatially distributed into individually addressable modes \cite{YumangNJP2019}. Through this process, the particle entanglement present in the initial state give origin to mode entanglement between its partitions \cite{KilloranPRL}, highlighting also a strong duality between these two concepts \cite{MatteoPRA2020}. After their first experimental realisation with Bose-Einstein condensates \cite{FadelScience2018}, split spin-squeezed states raised a lot of interest for their possible applications in quantum technologies and fundamental studies. Examples include theoretical investigations of their potential quantum metrology \cite{MatteoArXiv2022,jiajiePRAcondSS}, recently demonstrated experimentally in \cite{MaliaNat22}, and for investigating multipartite quantum correlations \cite{PRXidentical,PRLentquantBEC,BellBECByrnes,Vitagliano2023numberphase,FadelPRAnumberphase}. Taking this successful example into consideration, it would be crucial to understand whether other quantum information tasks could be accessible by such states. 

In this context, we provide here a new metrological protocol enabled by split spin-squeezed states. The idea is based on the fact that, due to the shared quantum correlations between the two parties of the system, performing a local measurement on one of them leaves the other in a conditional state that can have an extremely high sensitivity. This protocol can outperform the standard approach of using spin-squeezed states when the number of particles in the probe, as well as the state preparation time, are limited. 

Moreover, our measurement-based state preparation protocol can result in the generation of macroscopic superposition states, also known as spin Schr\"{o}dinger cat states \cite{MoelmerSoerensenGHZ,PawlowskiPRA2017}. Besides their interest for metrology, such states are appealing for fundamental studies of quantum correlations in many-body systems. Their non-classicality is notoriously related to interference fringes and negative regions in the Wigner function, which are typically difficult to prepare experimentally.

In summary, our work analyses a regime of system parameters and resources in which an assisted metrological protocol using split spin-squeezed states can offer an advantage. Moreover, we investigate the use of such states for the heralded preparations of macroscopic superposition states. These ideas could be implemented experimentally with Bose-Einstein condensates, where the preparation of cat-like spin states turned out extremely challenging using conventional approaches.

\section{Single probe metrology with OAT states}

In a typical quantum metrology scheme, the phase shift $\theta$ to be determined is encoded in a $N$-partite probe state $\rho_0$ by a generator $H$ as $\rho=e^{-i\theta H}\rho_0 e^{i\theta H}$. A fundamental limit to the maximum phase sensitivity is provided by the so-called quantum Cram$\acute{\text{e}}$r-Rao bound $\Delta \theta \geq \Delta \theta_{QCR} \equiv 1/\sqrt{v F_Q[\rho,H]}$, where $F_Q[\rho,H]$ is the quantum Fisher information (QFI) and $v$ is the number of independent measurements~\cite{PezzeRMP2018}. For a pure state, the QFI can be expressed in terms of the variance of $H$ as $F_Q[\rho,H]=4\var[\rho,H]$. The standard quantum limit tells us that for all classical states $F_Q[\rho,H]\leq N$, while according to the Heisenberg limit quantum states satisfy $F_Q[\rho,H] \leq N^2$. Therefore, observing $F_Q[\rho,H] > N$ implies the presence of metrologically useful entanglement \cite{PezzeRMP2018}. Moreover, a high QFI can be related to correlations that are even stronger than entanglement, namely Bell correlations \cite{PRAfisherbell}.

Of paramount importance for preparing atomic ensembles in quantum states with large QFI is the one-axis twisting (OAT) dynamics \cite{KitagawaPRA1993}. Starting from a $N$-partite spin coherent state pointing along the $+x$ direction, the OAT Hamiltonian $H=\hbar \chi S^2_z$ gives after an evolution time $t$ the state
\begin{align}\label{eq:SSS}
    \ket{\psi(\mu)} = \frac{1}{\sqrt{2^N}} \sum_{k=0}^N \sqrt{\binom{N}{k}} e^{-i\frac{\mu}{2}(N/2-k)^2}|k\rangle, 
\end{align}
where $\mu=2\chi t$ is an adimensional parameter, and $\ket{k}$ is the Dicke state with $k$ excitations. 

The properties of state Eq.~\eqref{eq:SSS} have been extensively investigated theoretically ~\cite{PezzeRMP2018,MAPR2011}. Notably, expectation values of the collective spin operator can be computed analytically, also for high moments \cite{JiajiePRL}.
This allows us to obtain analytical expressions also for the eigenvalues of the $3\times 3$ covariance matrix $\Gamma_{ij}=\text{Cov}[S_i,S_j]$, with $S_{i}\in\{S_x,S_y,S_z\}$. The basis change that diagonalizes $\Gamma$ is of clear physical intuition, and often convenient to use. Together with the polarization direction $x=x^\prime$, we introduce the squeezing direction $z^\prime = -\sin{\theta^*}y+\cos{\theta^*}z$ and anti-squeezing directions $y^\prime = \cos{\theta^*}y+\sin{\theta^*z}$, with
\begin{equation}
    \theta^* = \frac{1}{2} \arctan \left( \frac{4\sin{\left(\frac{\mu}{2}\right)} \cos^{N-2}{\left(\frac{\mu}{2}\right)}}{1-\cos^{N-2}(\mu)} \right),
\end{equation}
as the directions that respectively minimize ($z^\prime$) and maximize ($y^\prime$) the second moment of the collective spin.

The maximum eigenvalue of the covariance matrix $\Gamma$ is also proportional to the QFI of the state Eq.~\eqref{eq:SSS}. One obtains a QFI larger than $N$ for $0<\mu<2\pi$ and even reaching $N^2$ for $\mu=\pi$, when a ``Sch\"{o}dinger cat'' state is obtained \cite{MoelmerSoerensenGHZ}.

Experimentally, the OAT dynamics is implemented in e.g. ion traps through light-mediated interactions \cite{bohnetetal2016} or BECs through atomic elastic collisions \cite{RiedelNature2010}, and it is routinely used for the preparation of spin-squeezed states. These enabled numerous demonstrations of quantum-enhanced metrology, such as being applied to measuring magnetic fields \cite{OckeloenPRL2013}, improving frequency resolution in atomic clocks \cite{LudlowRMP2015,PedrozoNature2020}, and realizing squeezed matter-wave interferometry \cite{GreveNature2022}. 

If we consider a metrological application where the number of particles in the probe is limited to some maximum number, we also set a limit to the achievable QFI (\ie the Heisenberg limit), and thus to the sensitivity. However, one might argue that the state preparation could involve more particles than the one used in the probe itself, and ask whether this could be used to provide some advantage. While it is clear that if the ancillary particles are just discarded no advantage can be obtained, it is not trivial to see whether the probe sensitivity can be improved by a partial characterization of the ancillary particles' state. Here, by partial characterization we mean the information that can be obtained from some measurement of experimentally practical implementation, such as the result of a collective measurement performed on the ensemble of ancillary particles.

This question can be refined even further, by considering a more realistic situation that includes the relevant noise sources. In fact, during the preparation of squeezed BECs there are inevitable decoherence mechanisms resulting from technical and intrinsic noise~\cite{Li2009,YifanPRL2020}. The former can originate from imperfections in the implementation, while the latter is fundamental as it originates from particle losses. For BECs, these noise sources limit the OAT evolution to short times ($\mu < N^{-2/3}$).

\vspace{10mm}

\section{Assisted metrology}

In order to present our metrological protocol, we consider the case of an atomic ensemble in which the OAT dynamics is followed by a spatial separation of the particles into two distinct partitions~\cite{YumangNJP2019}, see Fig.~\ref{Fig1}. This last step can be realized by modifying the trapping potential to a double-well \cite{FadelBook}, or by exploiting additional internal states of the atoms \cite{KunkelPRL}, and it can formally be described by a beam-splitter transformation. The resulting split spin-squeezed state can thus be written as~\cite{YumangNJP2019} 
\begin{align}\label{eq:SSSS}
\ket{\Phi(\mu)} = \frac 1 {2^{N}} \sum_{N_A=0}^{N} \sum_{k_A=0}^{N_A} & \sum_{k_B=0}^{N_B} \sqrt{\binom{N}{N_A} \binom{N_A}{k_A} \binom{N_B}{k_B}} \nonumber \\
& \times e^{- i \frac{\mu}{2} (N/2 - k_A - k_B)^2} \ket{k_A}_{N_A} \ket{k_B}_{N_B} ,
\end{align}
where $N_\alpha$ is the number of particles for partition $\alpha \in \{A,B\}$ with $N_A+N_B=N$, and $\ket{k_\alpha}_{N_\alpha}$ is the $N_\alpha-$particle Dicke state with $k_\alpha$ excitations. Crucial to this state is that the multipartite entanglement generated by the OAT dynamics is partially ``converted'' by the spatial splitting into mode entanglement between the $A$ and $B$ partitions.

Split spin squeezed states have already been realized experimentally~\cite{FadelScience2018,PaoloPRX2023}, and are thus becoming relevant for practical metrological applications~\cite{MaliaNat22}. In the protocol we consider, $N_B$ particles constitute the probe, whose sensitivity might depends on the operations performed on the $N_A$ ancillary particles. In the following we investigate the probe's conditional states obtained upon performing a collective spin measurement on $A$, and discuss in which scenarios this assisted protocol can provide a better metrological performance than the standard OAT dynamics.

\subsection{Ideal scenario}

Let us consider the situation in which it is performed a measurement of the number of ancilla particles, and of their collective spin $S_{\vec{n}}^A$ along direction $\vec{n}$. Note that these two physical quantities can be measured simultaneously, as the associated operators commute. Obtaining as result $(N_A,l_A)$, the probe particles are left in the (unnormalized) state
\begin{equation}\label{eq:probeSt}
    \ket{\Phi(\mu)^B} = {}^{\vec{n}}_{N_A}\langle l_A \ket{\Phi(\mu)} ,
\end{equation}
where ${}^{\vec{n}}_{N_A}\langle l_A\vert$ is the $N_A$-particle Dicke state with $l_A$ excitations for $S_{\vec{n}}^A$. The probability for this to state to occur is given by
\begin{align}\label{eq:probeProb}
    p & (l_A,N_A|\vec{n}) =\frac{1}{2^{2N}} \sum_{k_A=0}^{N_A} \sum_{k_A^\prime=0}^{N_A} \sum_{k_B=0}^{N-N_A} \binom{N}{N_A} \binom{N-N_A}{k_B} \sqrt{\binom{N_A}{k_A} \binom{N_A}{k_A^\prime}} \nonumber \\
    & \times e^{-i\frac{\mu}{2}(N/2-k_A-k_B)^2} e^{i\frac{\mu}{2} (N/2-k_A^\prime-k_B)^2} \; {}^{\vec{n}}_{N_A}\langle l_A |k_A\rangle_{N_A} {}_{N_A}\langle k_A^\prime|l_A\rangle^{\vec{n}}_{N_A}  .
\end{align}
This expression allows us to introduce the probability of obtaining result $l_A$ from a measurement of $\hat{S}_{\vec{n}}^A$ on $N_A$ particles, namely $p_{N_A,\vec{n}}(l_A)=p(l_A,N_A|\vec{n})/p(N_A)$, where $p(N_\alpha) = 2^{-N} {{N}\choose{N_\alpha}}$ is the probability of having $N_\alpha$ particles in mode $\alpha\in\{A,B\}$.

For a given $N_A$ it is worth investigating the conditional states Eq.~\eqref{eq:probeSt}, their QFI, and their probability to occur Eq.~\eqref{eq:probeProb}. From our analytical expression it is possible to consider arbitrary measurement directions $\vec{n}$ and results $l_A$, but in the following we will focus on discussing the parameters we found most interesting. 

\begin{figure*}[t]
    \begin{center}
	\includegraphics[width=\textwidth]{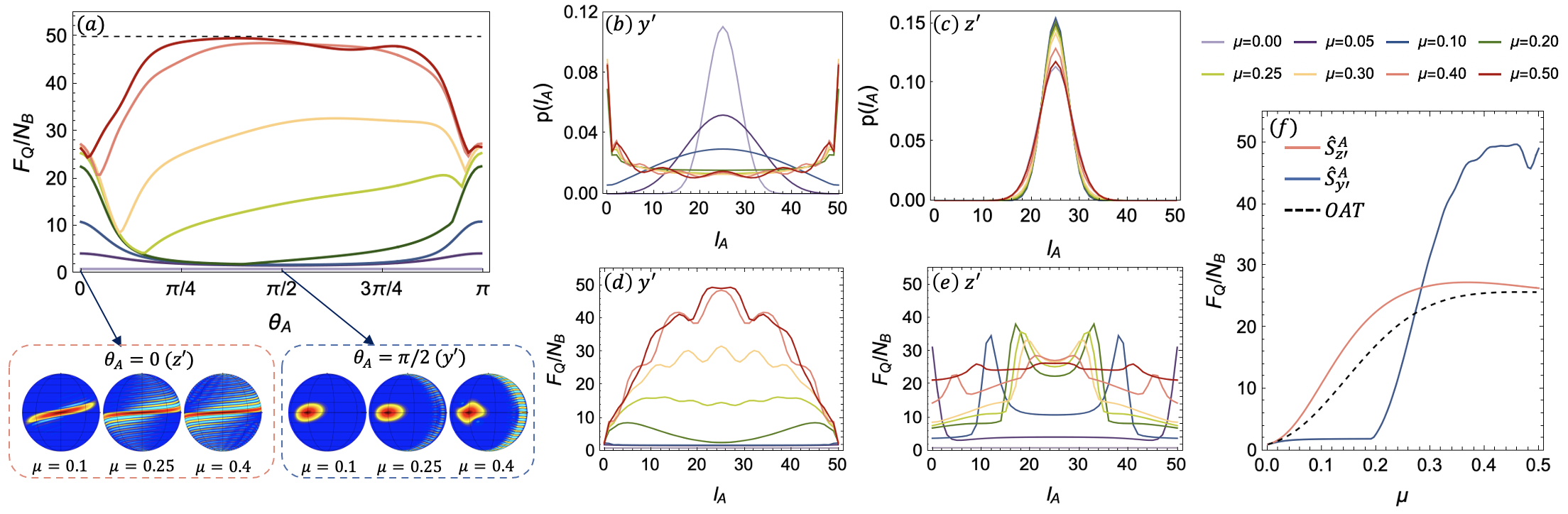}
	\end{center}
    \caption{\textbf{Measurements on the $\bf{xy}$-plane.} 
    Properties of the conditional states obtained from a split spin-squeezed state with $N=100, N_A=N_B=N/2$. 
    (a) For $l_A=N_A/2$, $F_Q/N_B$ as a function of the measurement direction $\theta_A$, and selected Wigner functions showing squeezed and cat-like states.
    Fixing $\theta_A$ such that the measurement direction is either $y'$ or $z'$, we show the probability of measuring $l_A$, panels (b,c) respectively, and the $F_Q/N_B$ of the associated states, panels (d,e). For $l_A=N_A/2$, we show in (f) a comparison between $F_Q/N_B$ of conditional states and of OAT states, as a function of the squeezing $\mu$. Here, the comparison is for a fixed number of particles $N_B$ in the probe state.
    }
    \label{Fig2}
\end{figure*}

We start considering a collective spin measurement on the $yz-$plane performed locally on $A$, so that $S_{\vec{n}}^A= \sin{\theta_A} S_{y^\prime}^A + \cos{\theta_A} S_{z^\prime}^A$, where $\theta_A$ is the angle between the measurement and the squeezing direction $z^\prime$. For $l_A=N_A/2$, we show in Fig.~\ref{Fig2}a the QFI of the conditional probe states as a function of $\theta_A$. Interestingly, for small values of $\mu$ conditional states with large QFI are obtained for $\theta_A=0$ (\ie the squeezing direction $z^\prime$), while for larger values of $\mu$ a large QFI is obtained for $\theta_A\approx\pi/2$ (\ie the antisqueezing direction $y'$). In order to understand better this behaviour, we look at the Wigner functions of the conditional probe states resulting from different measurement angles and levels of squeezing. Interestingly, we observe that a measurement along $\theta_A\approx 0$ results in conditional states that resemble spin squeezed and oversqueezed states, while a measurement along $\theta_A\approx\pi/2$ results in conditional states that resemble a superposition of coherent spin states, i.e. a spin cat state. To analyse the probability $p(l_A)\equiv p_{N_A,\vec{n}}(l_A)$ of these states to occur, we plot in Figs.~\ref{Fig2}b,c the value of Eq.~\eqref{eq:probeProb} for different levels of squeezing. As $\mu$ increases, if the measurement is performed along the anti-squeezing direction $y^\prime$, $p(l_A)$ tends to spread uniformly over all range of $l_A$ (see Fig.~\ref{Fig2}b), while for a measurement along the squeezing direction $z^\prime$, then $p(l_A)$ gets peaked around $l_A=N_A/2$ (see Fig.~\ref{Fig2}c). For both measurement directions, and for different results $l_A$, we can then compute the QFI of the conditional states, see Figs.~\ref{Fig2}d,e. 

With these in hand, we want to compare the metrological advantage given by the conditional probe states just investigated, and an OAT state. The resources we keep constrained are the number of atoms in the probe state, and the adimensional squeezing parameter $\mu$. For a nonlinearity $\chi$ independent of the particle number, the latter constraint corresponds to keeping fixed the state preparation time $t=\mu/2\chi$. In Fig.~\ref{Fig2}f we compare the value of $F_Q/N_B$ for the different conditional states just discussed, with the one for an OAT state with $N_B=50$ particles. This comparison is meaningful for a scenario where the number of particle in the probe is limited, but additional ancillary particles not interacting with the field to be estimated can be included in the state preparation and measurement. Interestingly, we see that there are situations where the conditional states reach much higher $F_Q/N_B$ than the OAT state, and that one can even saturate the Heisenberg limit, i.e. $F_Q/N_B\approx N_B$, for $\mu\ll \pi$.

In particular, when the measurement direction is aligned with the anti-squeezing direction $y^\prime$, we obtain for relatively large values of $\mu$ ($\mu > 0.4$ for $N_A=N_B=50$) conditional states with $F_Q/N_B$ that is in general high compared to a simple OAT state is observed. Interestingly, we also observe large fluctuations of $F_Q/N_B$ for $\mu>0.5$, and that it is possible to reach $F_Q/N_B\approx N_B$ for $\mu\ll \pi$ (see Supplementary Material~\cite{supplementary} Sec. \ref{SuppWigCond}). On the other hand, when the measurement is aligned with the squeezing direction $z^\prime$, the value of $F_Q/N_B$ obtained for the conditional states roughly follows the one for an OAT state, apart for small $\mu$. This regime is of particular interest, as i) these values of $\mu$ are the one typically explored in cold atom experiments, ii) in this regime one can exceed the $F_Q/N_B$ of an OAT state, and iii) this occurs with high probability, since $p(l_A)$ is peaked around $l_A=N_A/2$. Moreover, we will show in the following section that this configuration is also robust to noise, in the sense of particle number fluctuations and imperfect detection.

\begin{figure}
    \begin{center}
	\includegraphics[width=90mm]{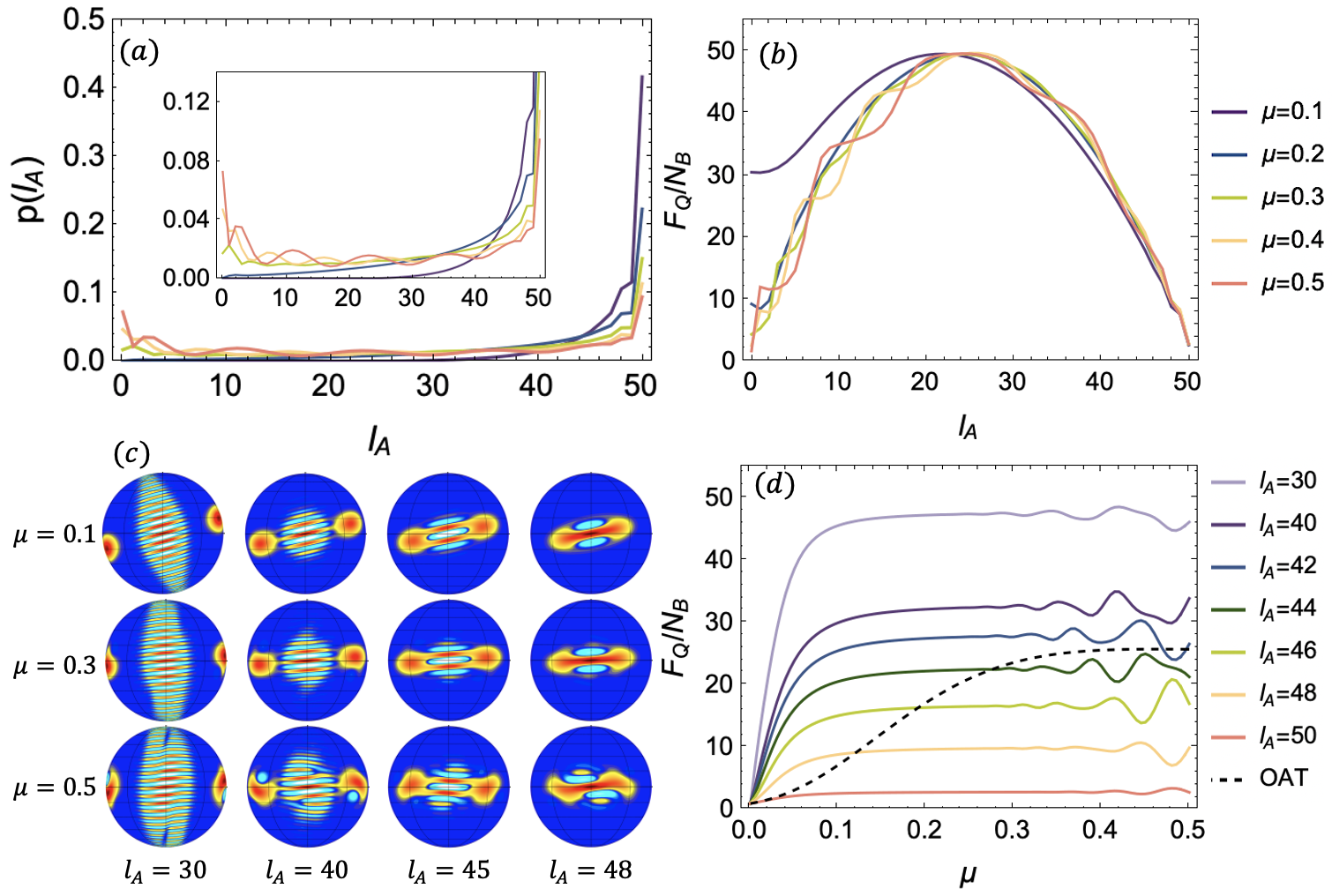}
	\end{center}
    \caption{\textbf{Measurements along the $\bf{x}$ direction.} 
    Properties of the conditional states obtained from a split spin-squeezed state with $N=100, N_A=N_B=N/2$. 
    (a) Probability of measuring $l_A$ for different levels of squeezing, and (b) the $F_Q/N_B$ of the associated conditional states. (c) Selected Wigner functions showing cat-like states of different size. (d) Comparison between  $F_Q/N_B$ of conditional states and of OAT states, as a function of the squeezing $\mu$. Here, the comparison is for a fixed number of particles $N_B$ in the probe state.
    }
    \label{figx}
\end{figure}

We then consider a collective spin measurement along $x$ performed locally on $A$. The probability to obtain a certain measurement result $l_A$ strongly depends on the amount of squeezing $\mu$, Fig.~\ref{figx}a. In fact, for $\mu=0$ the state is fully polarized along $x$, and one has $l_A=N_A$ with unit probability, but when $\mu$ increases the state starts to 'wrap around' the Bloch sphere, resulting in a non-zero probability for all possible $l_A$. The QFI for the associated conditional states is illustrated in Fig.~\ref{figx}b, showing a large variation even reaching the Heisenberg limit. If the result $l_A=N_A$ is obtained, the conditional probe state is a mildly squeezed spin state. However, as soon as one obtains $l_A<N_A$, the resulting conditional state resembles a spin cat state, Fig.~\ref{figx}c. Note that for $l_A>N_A/2$ the angular separation of the coherent spin states participating in the superposition, and therefore also the number of interference fringes, scales with $N_A-l_A$. Moreover, remember that even if conditional states with $F_Q/N_B\approx N_B$ are possible, these occur with very small probabilities. To compare the metrological advantage given by these conditional states and an OAT state we show in Fig.~\ref{figx}d the corresponding QFI values. For relatively large values of $\mu$ ($\mu > 0.3$ for $N_A=N_B=50$) we obtain conditional states with $F_Q/N_B$ that strongly fluctuates, taking values both larger and lower than the one of an OAT state. The behaviour is much more regular for small values of $\mu$, where we can see a regime in which conditional states with $l_A<N_A$ give a $F_Q/N_B$ growing in time much faster than the one of an OAT state (see $\mu<0.1$ in Fig.~\ref{figx}d). This regime is the one resulting in conditional states that closely resemble spin cat states.

In Sec. \ref{SuppSec1} of the Supplementary Material~\cite{supplementary} we give more details about the states considered so far, while in Sec. \ref{SuppWigCond} of ~\cite{supplementary} we show Wigner functions of the conditional states resulting from several other measurement directions and outcomes, together with their properties.

\subsection{Noisy scenarios}

So far we have considered a fixed $N_A$, but the splitting process resulting in the state Eq.~\eqref{eq:SSSS} is associated to partition noise which makes $N_A$ and $N_B=N-N_A$ fluctuate. For the equal (50:50) splitting we considered, the probability to observe $N_\alpha$ particles in mode $\alpha=A,B$ is simply given by the Binomial distribution $p(N_\alpha)$.
Concretely, this means that in an experiment the probe states will have a fluctuating number of particles and, therefore, a fluctuating sensitivity. 
In a practical scenario it would be extremely inefficient to post-select only experimental realisations with a given $N_B$, therefore we might ask what is the average sensitivity if all realisations are considered. In each realisation, $A$'s measurement gives knowledge of $N_A$ and $l_A$, which would allow us to perform a local optimisation on $B$ side to exploit the maximum sensitivity of the conditional state. We can thus define the average QFI density as
\begin{equation}\label{NBavQFI}
    \left\langle \dfrac{F_Q}{N_B}\right\rangle_{l_A} = \sum_{N_B=0}^N p(N_B) \dfrac{F_Q[\rho^B_{l_A,N_A|\vec{n}}]}{N_B} ,
\end{equation}
where $F_B[\rho^B_{l_A,N_A|\vec{n}}]$ is the QFI of the conditional probe state $\rho^B_{l_A,N_A|\vec{n}}$. The latter is obtained from a measurement on $A$ along the direction specified by $\vec{n}$, and giving as result $l_A$. However, note that $l_A$ has now to be a function of $N_A$, since the size of system $A$ is fluctuating. For example, we could compute Eq.~\eqref{NBavQFI} for the case when $l_A=\lceil N_A/2 \rceil$, which we have seen to be the most likely result for measurements on the $yz$ plane and small $\mu$, see Figs.~\ref{Fig2}b,c. Remarkably, we observe that there is no appreciable difference between a numerical evaluation of $\left\langle F_Q/N_B \right\rangle$ for measurements on the $yz$ plane and $l_A=\lceil N_A/2\rceil$, and the value of $F_Q/N_B$ when $N_A=N_B=N/2$ and $l_A=N_A/2$. In other words, averaging $F_Q/N_B$ over the distribution $p(N_B)$ seems to give a result compatible with the value of $F_Q/N_B$ when $N_B=N/2$. This could be explained by noting that: i) for large $N$ the distribution $p(N_B)$ is sharply peaked and symmetric around $N_B=N/2$, and ii) in the averaging, the $F_Q/N_B$ of a state with $N_B=N/2+k$ particles compensates the one of a state with $N_B=N/2-k$ particles, resulting in a value very close to the $F_Q/N_B$ of a state with $N_B=N/2$ particles. Perhaps surprisingly, we find that this correspondence holds for any value of $\mu$, and for different choices of the function for $l_A$ (e.g. $l_A=N_A-1$ for measurements of $S_x^A$).
More details about this comparison can be found in Sec. \ref{SI_noise} of Supplementary Material~\cite{supplementary}. There, we also compare another possible definition of average QFI, in the case where no measurement optimisation is done on $B$ side depending on the value of $N_B$ (still, the same post-selection according to $l_A(N_A)$ is applied). Even in this scenario, we observe that the average QFI is compatible with the value of $F_Q/N_B$ when $N_B=N/2$, which can be attributed to the fact that conditional states with different $N_B\approx N/2$ can appear very similar.

\vspace{3mm}

The second type of noise that we analyze is a measurement noise that results in errors on the observed value of $l_A$. Experimentally, this can originate from imperfect atom number counting, which always happens as detectors have finite resolution. We model this noise as a Gaussian distribution centered around $l_A$ and of standard deviation $\sigma$, such that if the value $l_A^\ast$ is observed there is a probability $p_{l_A,\sigma}(l_A^\ast)=(2 \pi \sigma^2)^{-1/2} e^{-(l_A^\ast-l_A)^2/2\sigma^2} $ for the true value to be $l_A$.

\begin{figure}
    \begin{center}
	\includegraphics[width=85mm]{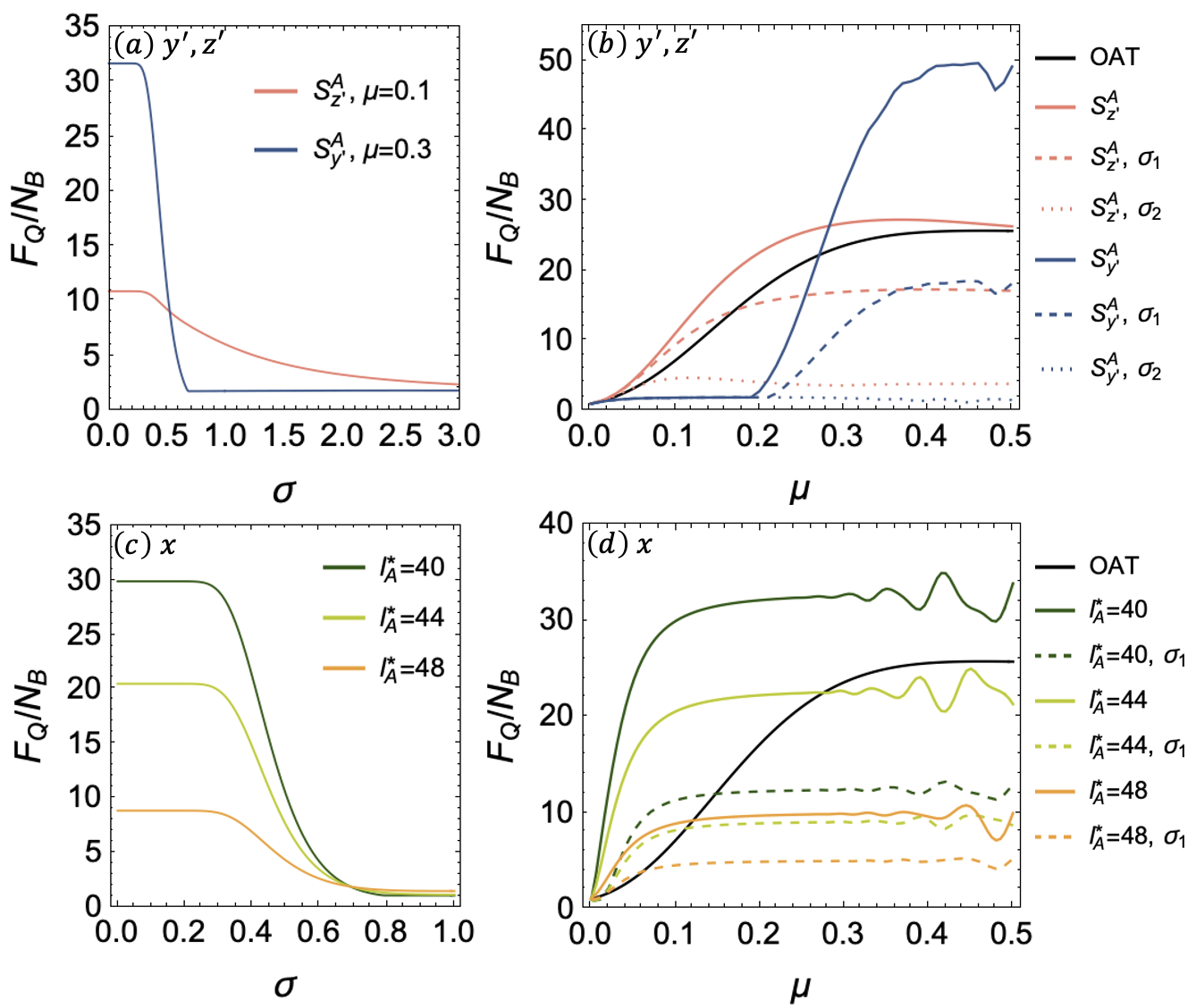}
	\end{center}
    \caption{\textbf{Robustness to detection noise.} 
    Sensitivity of the conditional states obtained from a split spin-squeezed state with $N=100, N_A=N_B=N/2$. Panels (a,b) consider conditional states with $l_A^\ast=N_A/2$ resulting from measurements along $y'$ and $z'$, while panels (c,d) along $x$. Panels (a,c) show $F_Q/N_B$ as a function of the detection noise $\sigma$, while (b,d) as a function of the squeezing $\mu$. Dashed lines are for a level of noise $\sigma_1=0.49$, while dotted lines for $\sigma_2=1.37$, which correspond to a $10\%$ probability of having $l_A=l_A^\ast\pm 1$ or $l_A=l_A^\ast\pm 2$ respectively.
    }
    \label{fig4}
\end{figure}

Analogously to the previous case, we define the QFI averaged over different values of $l_A$ as
\begin{equation}\label{lAavQFI}
    \left\langle \dfrac{F_Q}{N_B}\right\rangle_{l_A^\ast} = \dfrac{F_Q[\mathcal{N}\sum_{l_A=0}^{N_A} p_{N_A,\vec{n}}(l_A)p_{l_A,\sigma}(l_A^\ast) \rho^B_{l_A,N_A|\vec{n}}]}{N_B} ,
\end{equation}
where $\mathcal{N}^{-1}=\sum_{l_A=0}^{N_A} p_{N_A,\vec{n}}(l_A)p_{l_A,\sigma}(l_A^\ast)$ is a normalization parameter. We illustrate in Fig.~\ref{fig4} how this quantity varies as a function of $\sigma$, and for different measurement settings on $A$. 

Figure~\ref{fig4}a shows that the noise we are considering affects differently the conditional states obtained upon measurement of $S_{y^\prime}^A$ or $S_{z^\prime}^A$. In the first case, it appears that there exists a critical level of noise $\sigma^\star$ after which the average QFI is the one of a mixture of coherent spin states. On the other hand, this is not true in the second case, where we see the average QFI decreasing only asymptotically. We can understand this behaviour by looking at the Wigners in Fig.~\ref{Fig2}a, where it is reasonable to expect that the Gaussian-like conditional states resulting from $S_{z^\prime}^A$ measurements are more robust than the cat-like conditional states resulting from $S_{y^\prime}^A$ measurements. For a given amount of noise $\sigma$, it is interesting to know how the average QFI changes as a function of the squeezing $\mu$. This is illustrated in Fig.~\ref{fig4}b, for different levels of noise $\sigma$. Interestingly, while the fragility of the conditional states obtained from $S_{y^\prime}^A$ measurements results in an average QFI that can quickly fall below the value of the QFI for an OAT state, conditional states obtained from $S_{z^\prime}^A$ measurements seem able to achieve an average QFI larger than the one of an OTA state for small $\mu$, even if $\sigma$ is relatively large. This result further supports the statement made in the previous section, saying that the regime of small $\mu$ and $S_{z^\prime}^A$ measurements is of great interest for assisted metrology tasks, since it results in conditional states with high sensitivity and noise robustness.

Figure~\ref{fig4}c shows how the measurement noise we are considering affects the conditional states obtained upon measurement of $S_x^A$, for different values of the result $l_A^\ast$. As expected, the average QFI of conditional states with larger $l_A^\ast$ decays faster as the noise $\sigma$ increases, since such states are cat-like states with fine structures in the Wigner function that are rapidly washed-out by noise, see Fig.~\ref{figx}c. Also in this scenario, it is interesting to know how the average QFI changes as a function of the squeezing $\mu$, for a fixed amount of noise $\sigma$. This is illustrated in Fig.~\ref{fig4}d, for conditional states with different $l_A^\ast$. For small values of $\mu$ it is possible to see that the QFI of an OAT state can be surpassed by the considered conditional states, given a $l_A^\ast<N$ and a small enough $\sigma$. 

In addition, a further analysis on imperfection on atom counting on both $N_\alpha$ and $l_A$ is discussed in Supplementary Material ~\cite{supplementary} Sec. \ref{SI_NAnoise}.

\begin{figure}
    \begin{center}
	\includegraphics[width=70mm]{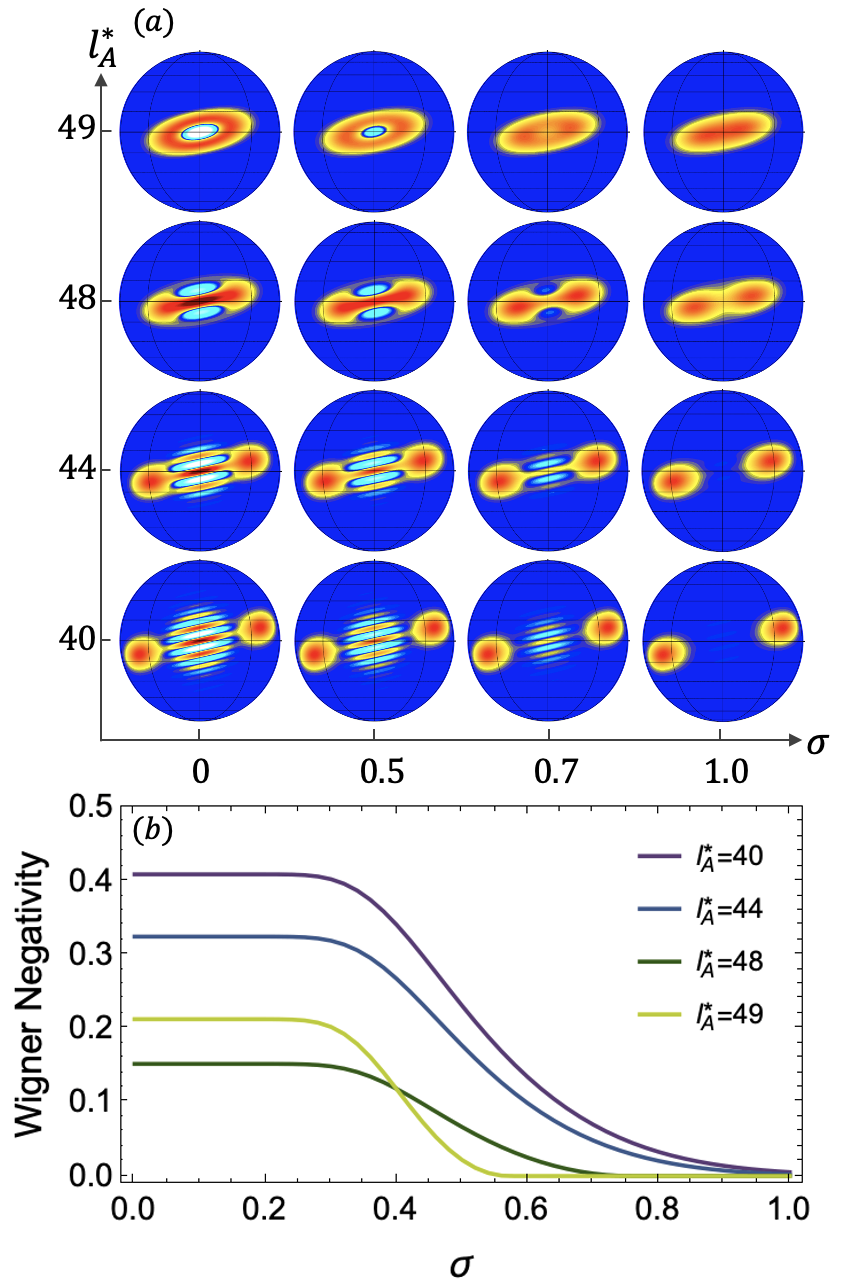}
	\end{center}
    \caption{\textbf{Heralded generation of cat states.} 
    Non-classical features of the conditional states obtained from a split spin-squeezed state with $N=100, N_A=N_B=N/2$, after measuring $S_x^A$. We show, as a function of the level of detection noise $\sigma$, (a) Wigner functions of conditional states associated to different $l_A^\ast$, and (b) Wigner function negativity as defined in Eq.\eqref{eqWN}.
    }
    \label{fig5}
\end{figure}

\section{Measurement-based preparation of spin cat states}

Sch\"{o}dinger cat states are regarded as powerful resources for quantum metrology, error-corrected quantum computing, and fundamental studies. While cat states have been successfully implemented with trapped ions \cite{MonroeScience1996,LoNature2015}, Rydberg atoms \cite{OmranScience2019},  optical and microwave photons \cite{AlexeiNature2007,HuangPRL2015,AuffevesPRL2013,SamuelNature2008,VlastakisScience2013}, and mechanical oscillators \cite{BildArXiv2022}, their realization atomic ensembles has remained elusive. Difficulties lie in engineering the correct nonlinear interactions, suppressing noise mechanisms (such as particle losses and phase noise), and performing measurements with high resolution. 

It is known that the OAT dynamics Eq.~\eqref{eq:SSS} result in a spin cat state at $\mu=\pi$ \cite{MoelmerSoerensenGHZ}. Nevertheless, following this simple strategy is unrealistic for BECs, due to the severe particle losses that would occur during the long dynamics. Approaches to mitigate these have been investigated \cite{PawlowskiPRA2017}, even if their experimental implementation remains challenging. Alternatively, ideas have been proposed to prepare macroscopic superpositions between two modes of a spin-1 BEC with a dynamic governed by spin-exchanging collisions \cite{PezzePRL2019}. 

In the analysis of conditional states we presented, we have seen that spin cat states can be obtained as a result of measurement along suitable directions (e.g. $S_{y^\prime}^A$ or $S_x^A$), if appropriate results are obtained, see Figs.~\ref{Fig2},\ref{figx}. We can thus propose to use this approach for the heralded preparation of macroscopic superposition states in spin-$1/2$ BECs. Crucially, even if this protocol demands a high-resolution in counting the number of particles, it has the advantage of being potentially fast, as the initial squeezed state that needs to be prepared requires an OAT evolution parameter $\mu$ much smaller than $\pi$ for $S_x^A$ measurements, see Fig.~\ref{figx}.

To understand the robustness of the protocol we propose, we investigate how finite measurement resolution affects the prepared state. From Figs.~\ref{figx},\ref{fig5} we can see that after a measurement of $S_x^A$ different cat states are obtained depending on the result $l_A$. In particular, these states have different size (i.e. separation between the two coherent spin state components), different parity (i.e. Wigner function value at the origin), and different orientation on the $yz$-plane. Therefore, if in an experiment the measured $l_A^\ast$ differs from the actual $l_A$ because of noise, the resulting conditional state will be a statistical mixture of different cat states. If this noise is too large, averaging over different cat states would result in a washing-out of the interference fringes, and thus of the quantum coherence of the superposition. Importantly, to estimate what is the amount of noise that can be tolerated it is not enough to take into account the variance $\sigma^2$ of the Gaussian distribution modeling uncertainties in $l_A$, but also the probability that a certain $l_A$ occurs for the parameters considered (See Eq.~\eqref{eq:plANA} of Supplementary Material~\cite{supplementary}). For this reason, if result $l_A^\ast$ is obtained, the conditional mixed state takes the form
\begin{equation}
    \rho(l_A^\ast,\sigma) = \mathcal{N}\sum_{l_A=0}^{N_A} p_{N_A,\hat{S}^A_{\vec{n}}}(l_A)p_{l_A,\sigma}(l_A^\ast) \rho^B_{l_A,N_A|\vec{n}} .
\end{equation}
In Fig.~\ref{fig5}a we plot the Wigner function of such states for different values of $l_A^\ast$ and $\sigma$. Even if the precise value of the noise that can be tolerated depends on the $l_A^\ast$ considered, we observe that around $\sigma\approx 0.7$ the interference fringes characterising the coherent superposition vanish. This value corresponds to an approximate probability of $p\approx0.2$ for the real value of $l_A$ to be $l_A^*\pm1$. 

A more quantitative analysis of the effect of noise is obtained by looking at the negativity of the Wigner function. For continuous variable systems, Wigner negativity is related to non-Gaussianity and non-classicality of the state \cite{KenfackJOB2004}, and it is known to be a resource for quantum information tasks \cite{TakagiPRA2018,AlbarelliPRA2018,MariPRL2012}. For spin systems, however, the Wigner function is defined on a (generalized Bloch) sphere, and the definition of non-Gaussianity and negativity is subtle. Here we follow Ref.~\cite{DavidPRR2021}, and compute the Wigner negativity as
\begin{equation}\label{eqWN}
    WN(\rho) = \frac{1}{2} \left(\frac{2j+1}{4\pi} \int_{\theta=0}^{\pi}\int_{\phi=0}^{2\pi} |W_{\rho}(\theta,\phi)|\sin\theta d\theta d\phi-1 \right),
\end{equation}
where $W_{\rho}(\theta,\phi)$ is the value of the Wigner function at point $(\theta,\phi)$ on the Bloch sphere, see Sec. \ref{SI_wig} of the Supplementary Material~\cite{supplementary}. In Fig.~\ref{fig5}b we show how the negativity of the conditional mixed states' Wigner function changes as a function of the amount of noise $\sigma$, for different values of $l_A^\ast$. For small $\sigma$, we have that $\rho$ stays very close to a pure state with $l_A^\ast\approx l_A$, so that its negativity stays constant until a critical value of $\sigma$ (approximately $0.4$ in Fig.~\ref{fig5}b) where conditional states with $l_A^\ast\pm 1$ start to contribute. After this point the negativity decreases until the point where it completely vanishes (approximately $0.5-0.9$ in Fig.~\ref{fig5}b).

\section{Conclusions}
We analysed the conditional states resulting from a local measurement in one of the two parts of a split-spin squeezed state. The multipartite entanglement present in these states, combined with the local measurement, leads to a rich family of non-trivial conditional states exhibiting high Fisher information or large Wigner negativities. These have been investigated quantitatively, for different local measurement directions and outcomes, both without and with the presence of noise. The latter was chosen to take into account particle number fluctuations in the conditional states, which are intrinsic in the probabilistic (beam-splitter-like) splitting process, as well as measurement imperfections. We observe that the observed non-classical properties are robust to noise, and therefore of interest for applications in quantum technologies.

In this context, we propose a protocol that can be used to enhance the sensitivity of a measurement probe in a scenario where its size, as well as the state preparation time, are limited. Our idea is based on the fact that, if the probe is entangled with an ancilla system, a local measurement in the latter can prepare the probe in conditional states with much higher sensitivity. Concretely, we analyse a scenario where a split spin-squeezed state is shared between the probe and the ancilla, and identified the range of system parameters and local measurements providing a metrological advantage.

Besides this practical application, we note that the measurement-based state preparation protocol we investigate can be used to generate spin cat states. These macroscopic superposition states are of interest not only for metrology, but also for fundamental research. We quantify the non-classicality of the conditional states that can be prepared through a measure of their Wigner function negativity, and investigated its robustness with noise.   

A natural platform where our ideas could be realized are ultracold atomic ensembles, where spin-squeezed states are routinely prepared for a number of applications. More recently, the spatial splitting of such states was also demonstrated \cite{FadelScience2018,PaoloPRX2023}, thus opening the path to the experimental study of split spin-squeezed states \cite{YumangNJP2019}. Apart from shedding light on multipartite quantum correlations \cite{PRXidentical,PRLentquantBEC,BellBECByrnes}, it is of interest to investigate the usefulness of such states for quantum technologies, such as for quantum teleportation \cite{ManishPRA2021}, and metrology \cite{MatteoArXiv2022,jiajiePRAcondSS}. Our study brings a contribution in these interesting directions.

\vspace{6mm}
\textbf{Acknowledgements.--} This work is supported by the National Natural Science Foundation of China (Grants No. 12125402, No.11975026 and  No.12147148) and the Beijing Natural Science Foundation (Z190005). JG acknowledges financial support from the China Scholarship Council (Grant No. 202106010192). FS acknowledges the China Postdoctoral Science Foundation (Grant No. 2020M680186). MF was supported by The Branco Weiss Fellowship -- Society in Science, administered by the ETH Z\"{u}rich.

\bibliographystyle{apsrev4-1} 
\bibliography{CatReference.bib}

\clearpage
\newpage

\appendix

\begin{widetext}
\section*{Supplementary information}

\subsection{One-axis twisting dynamics and and Quantum Fisher Information}\label{SuppSec1}

\subsubsection{Spin-squeezed and split spin-squeezed states}
The dynamics resulting from a one-axis twisting (OAT) dynamics has been analyzed extensively in the literature, so we will just have here a short discussion to complement the main text.
Starting from a spin coherent state polarized along the $x$ direction, the (OAT) Hamiltonian $H=\hbar\chi S_z^2$ gives after a time $t=\mu/2\chi$ the state
\begin{equation}\label{eq:appSSS}
\ket{\psi(\mu)} = \dfrac{1}{\sqrt{2^N}} \sum_{k=0}^N \sqrt{{N}\choose{k}} e^{- i \frac{\mu}{2} (N/2 - k)^2} \ket{k},
\end{equation} 
where $\ket{k}$ denotes the $N$-qubit Dicke state with $k$ excitations.

For short evolutions $\mu\approx 0$, the OAT evolution results in spin-squeezed states. On the other hand, longer evolution times result in highly non-Gaussian states, such as spin-oversqueezed or cat states, see Fig.~\ref{SI_S1fig1}a. These states are in general very sensitive to rotations, as it can be seen from their high quantum Fisher information, Fig.~\ref{SI_S1fig1}b.
\begin{figure*}[h]
	\begin{center}
		\includegraphics[width=120mm]{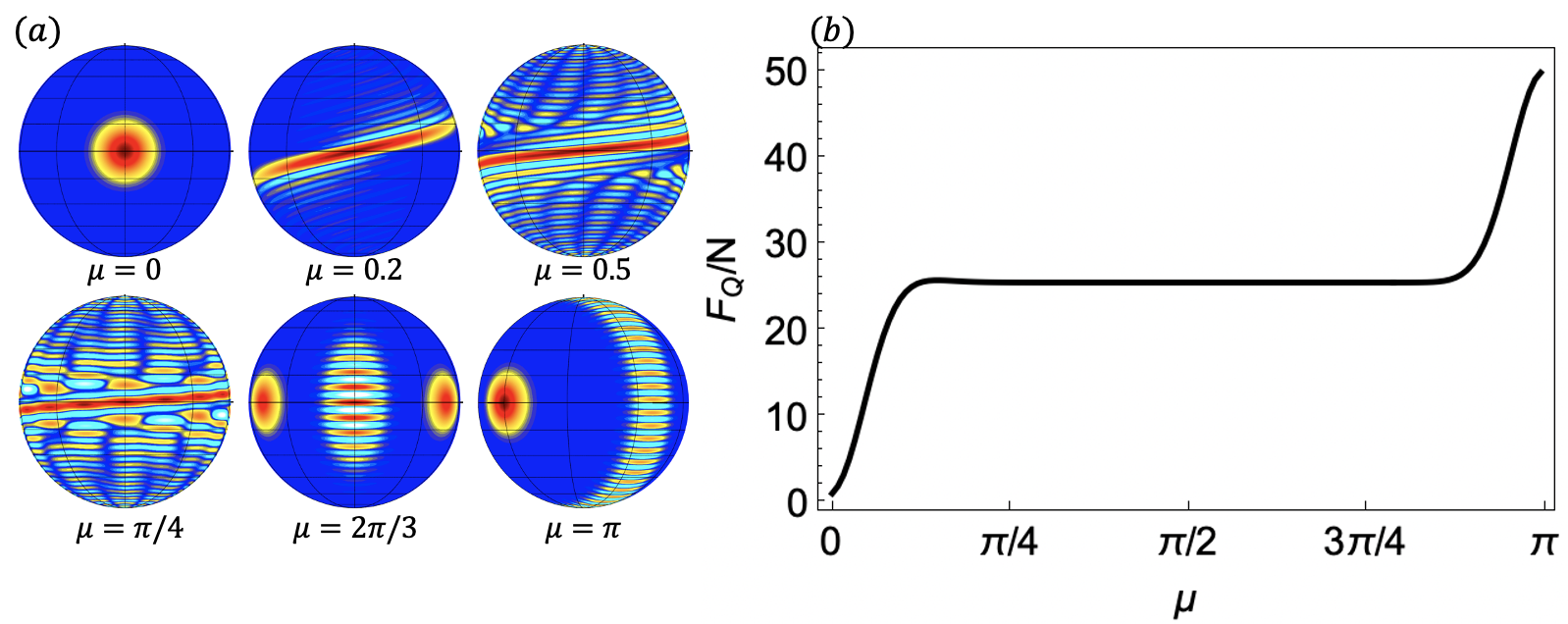}
	\end{center}
	\caption{\textbf{One-axis twisted states.} 
    For $N=50$, (a) selected Wigner functions of the OAT states obtained for different $\mu$, and (b) $F_Q/N$ as a function of $\mu$.
    }
	\label{SI_S1fig1}
\end{figure*}

If the particles in state Eq.~\eqref{eq:appSSS} are spatially separated into two distinct regions, to form ensembles $A$ and $B$, under the assumption that the splitting process is ``beam-splitter-like'' we obtain the state \cite{YumangNJP2019}
\begin{align}\label{SIeqSSS}
\ket{\Phi(\mu)} = \frac 1 {2^{N}} \sum_{N_A=0}^{N}\sum_{k_A=0}^{N_A}\sum_{k_B=0}^{N-N_A}  \sqrt{\binom{N}{N_A} \binom{N_A}{k_A} \binom{N-N_A}{k_B}} e^{- i \frac{\mu}{2} (N/2 - k_A - k_B)^2} \ket{k_A}_{N_A} \ket{k_B}_{N-N_A} \;.
\end{align}
Here, $\ket{k}_{N}$ indicates the $N$-qubit Dicke states with $k$ excitations, and we have assumed a $50:50$ splitting ratio. Analytical expressions for arbitrary splitting ratios, and for splitting into more than two modes, can also be derived analytically.

In the assisted metrological protocol we consider, one performs on the ancilla particles a measurement of the local collective spin
\begin{align}
S_{\vec{n}}^A=\sum_{N_A=0}^N\sum_{l_A=0}^{N_A}\left(l_A-\frac{N_A}{2}\right)\ket{l_A}_{\mathbf{n},N_A}\bra{l_A}_{\mathbf{n},N_A} ,
\end{align}
where $\vec{n}$ is a unit vector in $\mathbb{R}^3$ specifying the spin projection axis, and $\ket{l}_{\mathbf{n},N}$ the associated eigenstates. The latter can be seen as $N$-qubit Dicke states in the $\vec{n}$-basis with $l$ excitations.

Measuring $S_{\vec{n}}^A$ gives as result the values $\{N_A,l_A\}$, and projects the probe ($B$'s) state is a conditional state. This can be calculated by applying the projector $\ket{l_A}_{\mathbf{n},N_A}\bra{l_A}_{\mathbf{n},N_A}$ to Eq.~\eqref{SIeqSSS}, which gives 
\begin{align}\label{eq:condstate}
\ket{\Phi(\mu)^B}_{l_A,N_A|\vec{n}}=\frac {p(l_A,N_A|\vec{n})^{-1/2}} {2^{N}} \sum_{k_A=0}^{N_A}\sum_{k_B=0}^{N-N_A}\sqrt{\binom{N}{N_A}\binom{N_A}{k_A}\binom{N-N_A}{k_B}} e^{- i \frac{\mu}{2} (N/2 - k_A - k_B)^2} \bra{l_A}_{\vec{n},N_A}\ket{k_A}_{N_A} \ket{k_B}_{N-N_A} ,
\end{align}
where $p(l_A,N_A|\vec{n})$ is the probability to obtain $\{N_A,l_A\}$ from a measurement of $S^A_{\textbf{n}}$, namely 
\begin{align}\label{eq:plANA}
p(l_A,N_A|\vec{n})=\frac 1 {2^{2N}} \sum_{k_A=0}^{N_A}\sum_{k'_A=0}^{N_A}\sum_{k_B=0}^{N-N_A}\binom{N}{N_A}\binom{N-N_A}{k_B}\sqrt{\binom{N_A}{k_A}\binom{N_A}{k'_A}} e^{- i \frac{\mu}{2} (N/2 - k_A - k_B)^2}e^{i \frac{\mu}{2} (N/2 - k'_A - k_B)^2} \bra{l_A}_{\vec{n},N_A}\ket{k_A}_{N_A}\bra{k'_A}_{N_A}\ket{l_A}_{\vec{n},N_A} .
\end{align}

These expressions allow us to compute sensitivities and heralding probabilities of the conditional states.

\subsubsection{Quantum Fisher information}
The quantum Fisher information (QFI) quantifies the sensitivity of a quantum state $\rho$, to a perturbation generated by an operator $H$. It can be computed as
\begin{align}
    F_Q[\rho,H] = 2\sum_{\kappa,\kappa' q_{\kappa}+q_{\kappa'}>0} \frac{(q_{\kappa}-q_{\kappa'})^2}{q_{\kappa}+q_{\kappa'}} |\langle \kappa'|H|\kappa \rangle|^2 ,
\end{align}
where $q_{\kappa}$ are the eigenvalues of $\rho$, and $\kappa$ their associated eigenvectors.

For arbitrary quantum states it holds the inequality $F_Q[\rho,H]\leq 4\text{Var}[\rho,H]$, which is saturated for pure states $\rho=|\psi \rangle \langle \psi|$. This allows us to compute $F_Q$ from the maximum eigenvalue of the covariance matrix for the state $|\psi\rangle$. Namely, we consider $H$ to be in the space generated by $\{S_x,S_y,S_z\}$, and compute the $3\times 3$ covariance matrix with elements $[\Gamma]_{ij}=\text{Cov}[A,B]$, with $A,B\in \{S_x,S_y,S_z\}$. Then, $F_Q=\lambda_{\text{max}} (\Gamma)$, where the eigenvector associated to the maximum eigenvalue $\lambda_{\text{max}}$ of $\Gamma$ specifies the most sensitive rotation axis, and thus the optimal generator of rotations $H^\ast$.

In several cases, however, we need to calculate the QFI of a mixed state. We thus define the generalized $3\times3$ matrix \cite{PezzeRMP2018}
\begin{align}
    [\Gamma_Q]_{ij} = 2 \sum_{\kappa,\kappa' q_{\kappa}+q_{\kappa'}>0} \frac{(q_{\kappa}-q_{\kappa'})^2}{q_{\kappa}+q_{\kappa'}} \langle \kappa'|J_i|\kappa\rangle \langle \kappa | J_j |\kappa'\rangle ,
\end{align}
from which we can compute the QFI as $F_Q=\lambda_{max} (\Gamma_Q)$.

\clearpage
\newpage

\subsection{Detailed analysis of conditional states}\label{SuppWigCond}

\subsubsection{Wigner function plots of $B$'s conditional states}

\begin{figure}[t]
	\begin{center}
		\includegraphics[width=140mm]{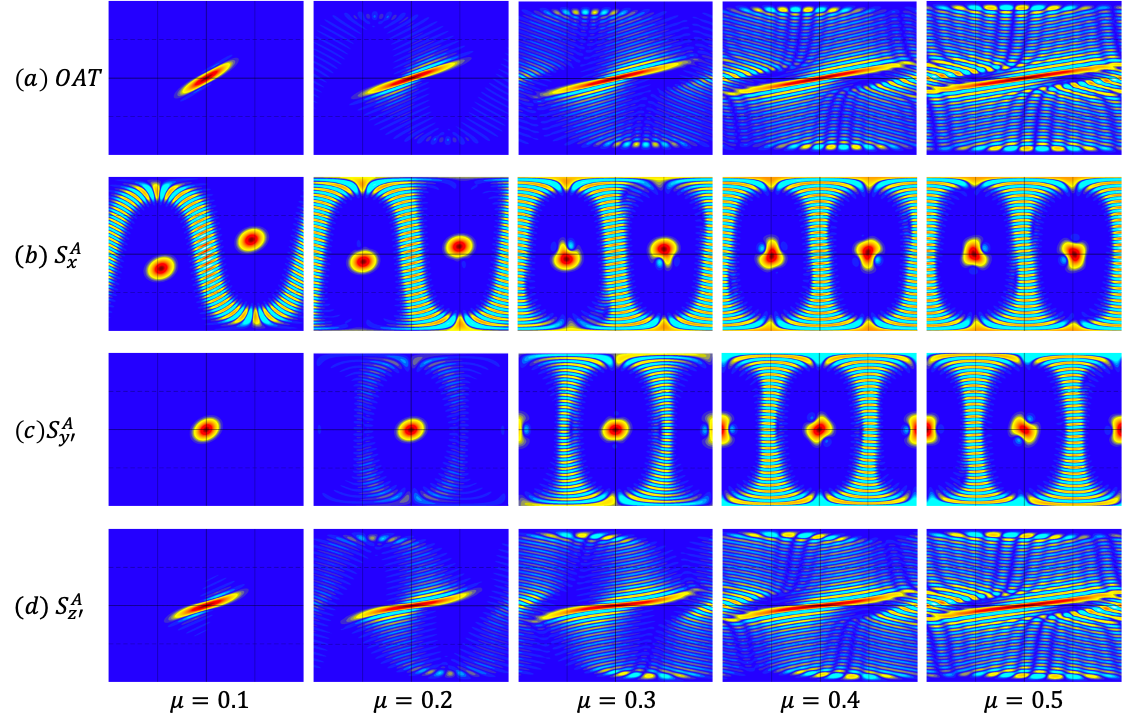}
	\end{center}
	\caption{\textbf{Wigner function comparison I.} 
    Selected Wigner functions of OAT states with $N=50$, panel (a), compared to Wigner functions of conditional states obtained after measuring a $N=2N_A=100$ split spin-squeezed state along different directions (b-d).
    }
	\label{SI_S2fig1}
\end{figure}

\begin{figure}
    \begin{center}
	\includegraphics[width=130mm]{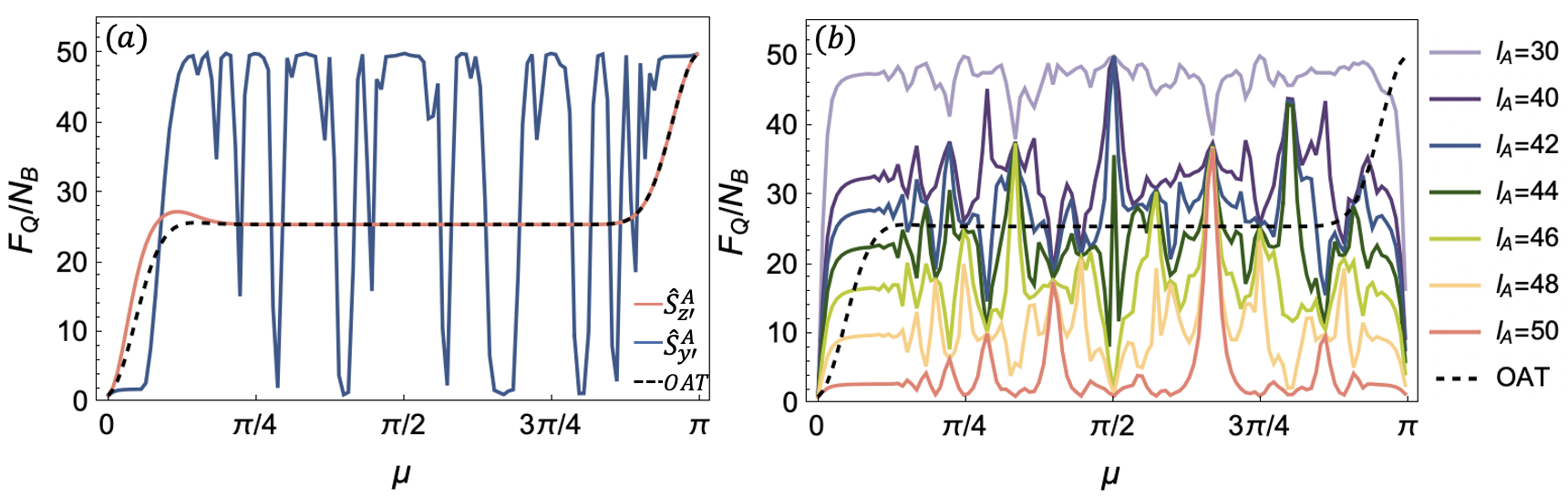}
	\end{center}
    \caption{\textbf{Conditional states' sensitivity.} 
    Considering a split spin-squeezed state with $N=100, N_A=N_B=N/2$, value of $F_Q/N_B$ for conditional states obtained from measuring $y'$ or $z'$, panel (a), or $x$, panel (b), as a function of the squeezing $\mu$.	These are compared to the $F_Q/N_B$ of a OAT state with $N=50$ (black dashed line).
    }
    \label{SIfig1b}
\end{figure}

\begin{figure*}[h]
	\begin{center}
		\includegraphics[width=140mm]{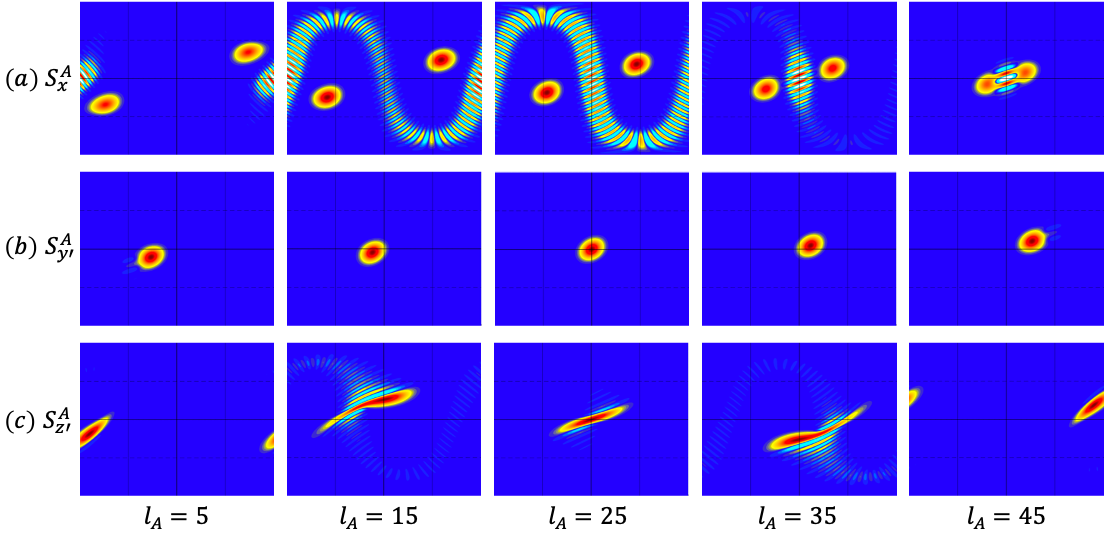}
	\end{center}
	\caption{\textbf{Wigner function comparison II.} 
    Selected Wigner functions of conditional states resulting by measuring a $N=2N_A=100$ split spin-squeezed state along different directions, as a function of the result obtained $l_A$.
    }
	\label{SI_S2fig2}
\end{figure*}
In Figure~\ref{SI_S2fig1} we compare the Wigner functions of an OAT state~\eqref{eq:appSSS} with the ones of $B$'s conditional states~\eqref{SIeqSSS}. At small squeezing level $\mu$, the OAT dynamics can only produce spin squeezed and over-squeezed states. However, for the same evolution time, cat states can be obtained from our measurement-assisted protocol after a measurement of $S_x^A$ or $S_{y^\prime}^A$ is performed on $A$. If instead $S_{z^\prime}^A$ is measured, the resulting states will be very similar to the OAT spin states. 

In Fig.~\ref{SIfig1b} we illustrate the QFI of $B$'s conditional states for $l_A$ fixed, but different $\mu$. Contrary to Fig.~2f and Fig.~3d of the main text, where we plot the same curves but for a smaller range of $\mu$ that is of interest, here we show the QFI over a full period of OAT dynamics. As mentioned in the main text, one can observe that for measurements of $S_x^A$ or $S_{y^\prime}^A$, the value of $F_Q/N_B$ strongly fluctuates, taking values both larger and lower than the one of an OAT state. On the other hand, for measurements of $S_{z^\prime}^A$ the value of $F_Q/N_B$ mostly follows the one of an OAT state, except for small squeezing values.

Finally, we investigate how the conditional states depend on the measurement outcome $l_A$. Representative Wigner functions are shown in Fig.~\ref{SI_S2fig2}. For a measurement of $S_x^A$, we can see that different $l_A$ result in cat states of different size, while for a measurement of $S_{y^\prime}^A$, $S_{z^\prime}^A$, we can see that there is a shift in the position of the state depending on $l_A$.

Here note that, if a measurement of $S_z^A$ is performed on $A$, according to Eq.~(\ref{eq:condstate}) the conditional states will simply be
\begin{align}\label{eq:SzAcondB}
    |\Phi(\mu)^B\rangle_{l_A,N_A|S_z^A} = \frac{1}{2^{N-N_A}}\sum_{k_B=0}^{N-N_A} \sqrt{\binom{N-N_A}{k_B}} e^{-i\frac{\mu}{2}(N/2-l_A-k_B)^2}|k_B\rangle_{N-N_A} ,
\end{align}
namely spin squeezed states rotated around the $z$-axis by an angle proportional to $l_A$.

\subsubsection{Sensitivity in noisy scenarios}\label{SI_noise}

In Fig.~\ref{figNBfluct}a we illustrate the QFI for conditional states with different number of particles $N_B$ (with $N_A+N_B=N$ fixed). Since the splitting process results with high probability in $N_B\in[2N/5,3N/5]$, we only restrict to plotting a few lines for clarity. As one expects, states with larger $N_B$ result in larger QFI. However, since $N_B$ fluctuates symmetrically around its mean value $N/2$, we observe that fluctuations in QFI cancel out, in the sense that the average QFI as defined in Eq.~(7) of the main text is indistinguishable from the one of a state with fixed $N_B=N/2$ particles. This is shown in Fig.\ref{figNBfluct}b, where the curves for each case considered completely overlap. In addition, we plot also the value of the average QFI defined as
\begin{equation}\label{eq:aveQFINB2}
    F_Q(\overline{\rho^B_{l_A,N_A}})/\overline{N_B} = \dfrac{F_Q[\oplus_{N_B=0}^N p(N_B) \rho^B\vert_{N_A,l_A}]}{\sum_{N_B=0}^N p(N_B)N_B} \;,
\end{equation}
which we see to also be indistinguishable from the one of a state with fixed $N_B=N/2$ particles.

\subsection{Noises on local atom number} \label{SI_NAnoise}
In the main text, we consider the case of $N_A=N/2$, which is the most likely event. Due to the effect of beam splitter, the atom number for both sides is not a certain number, but in a binomial distribution:
\begin{align}
p(N_{\alpha}) = \frac{1}{2^N} \binom{N}{N_\alpha}.
\end{align}
With detection noise on $l_A$, the average QFI of Bob's conditional states over all possible $N_A$ is
\begin{align}\label{QFINAave}
\left\langle \dfrac{F_Q}{N_B}\right\rangle = \dfrac{ F_Q[ \mathcal{N}\bigoplus_{N_A=0}^{N}p(N_A) \sum_{l_A=0}^{N_A} p_{N_A,\vec{n}}(l_A)p_{l_A,\sigma}(l_A^\ast) \rho^B_{l_A,N_A|\vec{n}}]}{\sum_{N_B=0}^Np(N_B)N_B},
\end{align}
where $\mathcal{N}$ is the normalization parameter. We compare the condition where QFI average over all possible $N_A$ in Eq.~(\ref{QFINAave}) to the one with $N_A=N/2$ in Eq.~(7) of the main text in Fig.~\ref{figNAnoise}, and the difference between them is quite small. That is because in our model the fluctuation of $N_A$ only increases the dimension of Hilbert space on Bob's conditional states, but has negligible effect on decoherence.

Besides measurement noise on $l_A$, the imperfection on atom counting will also affect on $N_A$. So here, we consider the same detection noisy scenarios on $N_A$. The noise is a Gaussian distribution around $N_A$ and with the same standard deviation $\sigma$ as $l_A$ detection noise. If the value $N_A^*$ is observed, there is a probability $p_{N_A,\sigma}(N_A^*)=(2\pi \sigma^2)^{-1/2}e^{-(N_A^*-N_A)^2/2\sigma^2}$ for the true value to be $N_A$. After taking detection noises on both $l_A$ and $N_A$, we define the resulting QFI as
\begin{align}\label{QFINAnoise}
\left\langle \dfrac{F_Q}{N_B}\right\rangle_{N_A^*, l_A^*} = \dfrac{ F_Q[ \mathcal{N}\bigoplus_{N_A=0}^{N}p(N_A)p_{N_A,\sigma}(N_A^*) \sum_{l_A=0}^{N_A} p_{N_A,\vec{n}}(l_A)p_{l_A,\sigma}(l_A^\ast) \rho^B_{l_A,N_A|\vec{n}}]}{\sum_{N_B=0}^Np(N_B)N_B},
\end{align}
with normalization parameter $\mathcal{N}$. In Fig.~\ref{figNAnoise}, we take detection noise on $N_A$ into account by simulating the noise in a Gaussian distribution. We find that the fluctuation on $N_A$ leads to negligible effect on sensitivity.

\begin{figure}
    \begin{center}
	\includegraphics[width=12cm]{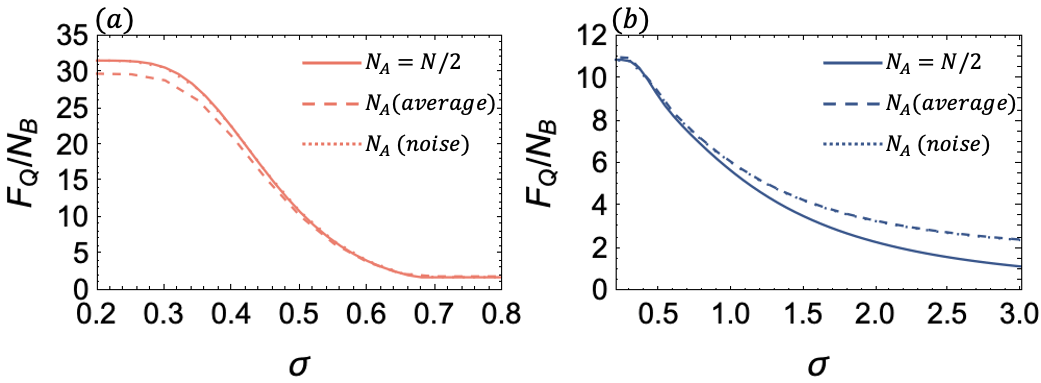}
	\end{center}
    \caption{\textbf{Effect of noise on atom number $N_A$.} 
    Sensitivity of the conditional states obtained from a split spin-squeezed state with $N=100$, with (a) $\mu=0.3$ and Alice measures $S_{y'}^A$ (anti-squeezing direction), (b)$\mu=0.1$ and Alice measures $S_{z'}^A$ (squeezing direction). The measurement noise on $l_A$ is always considered here. We compare $N_A=N/2$ in Eq.~(7) of the main text (solid lines), the average over $N_A$ in Eq.~(\ref{QFINAave}) (dashed lines) and detection noise on $N_A$ in Eq.~(\ref{QFINAnoise}) (dotted lines), and we find that the fluctuation on $N_A$ leads to negligible effect on sensitivity. }
    \label{figNAnoise}
\end{figure}

\subsection{Wigner negativity for spin states}\label{SI_wig}
The Wigner function associated with the polar angle $\theta\in[0,\pi]$ and the azimuthal angle $\phi\in[0,2\pi]$ in phase space is defined as~\cite{DowlingPRAWig,TimBook2020}
\begin{align}
    W(\theta,\phi) = \sum_{l=0}^{2j} \sum_{m=-l}^{l} \rho_{lm} Y_{lm} (\theta,\phi) ,
\end{align}
where $Y_{lm} (\theta,\phi)$ are the spherical harmonic functions, and $\rho_{lm}$ is the matrix element defined
\begin{align}
    \rho_{lm}=\sum_{m_1=-j}^{j} \sum_{m_2=-j}^{j} (-1)^{j-m_1-m} \langle j m_1;j -m_2| lm \rangle \langle j m_1|\rho|j m_2\rangle ,
\end{align}
with $\langle j_1 m_1;j_2 m_2| lm \rangle$ the Clebsch-Gordan coefficient. And since the these coefficients are zero unless $m=m_1-m_2$, the Wigner function can be written as
\begin{align}
    W_{\rho}(\theta,\phi)=\sum_{l=0}^{2j} \sum_{m_1=-j}^{j} \sum_{m_2=-j}^{j} (-1)^{j-2m_1+m_2} \langle j m_1;j-m_2| lm_1-m_2 \rangle Y_{lm_1-m_2}(\theta,\phi)  \langle j m_1 |\rho |j m_2\rangle .
\end{align}
According to~\cite{DavidPRR2021}, the Wigner negativity (WN) of a spin state can be calculated from this expression as
\begin{align}
    WN(\rho)=\frac{1}{2} \left(\frac{2j+1}{4\pi} \int_{\theta=0}^{\pi}\int_{\phi=0}^{2\pi} |W_{\rho}(\theta,\phi)|\sin\theta d\theta d\phi-1 \right).
\end{align}
In Fig.~5 of the main text we plot the value of WN for the conditional states we considered, and show that is non-zero for reasonable values of detection noise.

\begin{figure}
    \begin{center}
	\includegraphics[width=\textwidth]{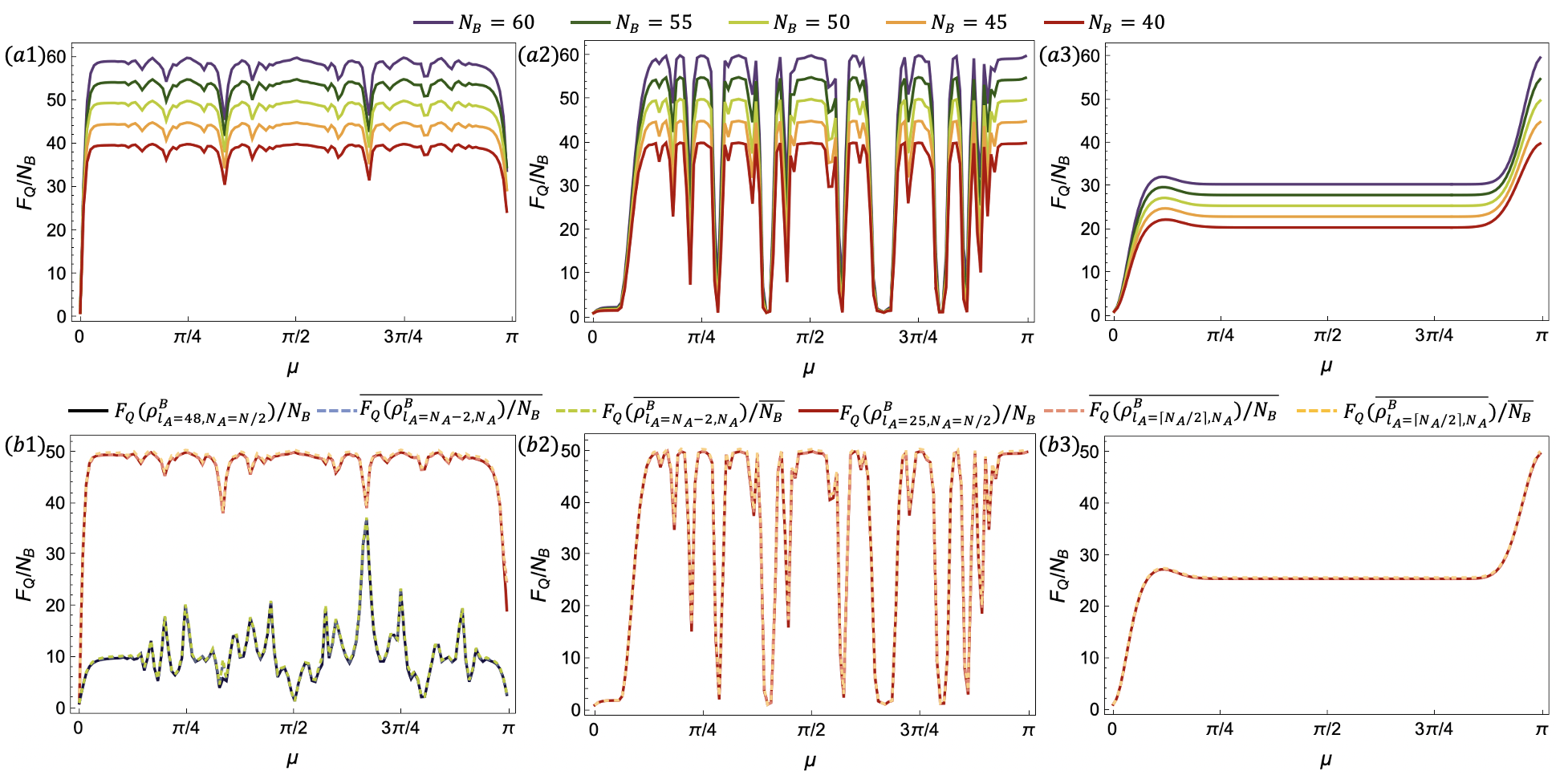}
	\end{center}
    \caption{\textbf{Effect of noise on the conditional states' sensitivity.} 
    Panels $(a)$ show $F_Q/N_B$ for conditional states with different $N_B$ (but fixed $N=100$), as a function of $\mu$ and for different measurement directions. Panels (b) show that the average $\overline{F_Q(\rho^B_{l_A,N_A})/N_B}$ given in Eq.~(6) of the main text, and $F_Q(\overline{\rho^B_{l_A,N_A}})/\overline{N_B}$ given in Eq.~\eqref{eq:aveQFINB2}, are effectively equivalent to $F_Q(\rho^B_{l_A,N_A=N/2})/N_B$.}
    \label{figNBfluct}
\end{figure}

\end{widetext}

\end{document}